\documentclass{article}
\usepackage{graphicx,amsmath,amssymb,a4wide}

\newtheorem{theorem}{Theorem}

\begin{document}
\begin{center}

{\huge Diffraction of a binary non-Pisot inflation tiling}\vspace{4ex}

{\Large Michael Baake$^{1,3}$ and Uwe Grimm$^{2,3}$}\vspace{4ex}

{\small
$^1$Fakult\"{a}t f\"{u}r Mathematik, Universit\"{a}t Bielefeld,
Postfach 100131, 33501 Bielefeld, Germany\\[0.25ex]
$^2$School of Mathematics and Statistics, The Open University,
Milton Keynes MK7 6AA, UK\\[0.25ex]
$^3$School of Mathematics and Physics, University of Tasmania,
Hobart TAS 7001, Australia\\[0.25ex]}\vspace{2ex}
\end{center}

\begin{abstract}
  A one-parameter family of binary inflation rules in one dimension is
  considered. Apart from the first member, which is the well-known
  Fibonacci rule, no inflation factor is a unit. We identify all cases
  with pure point spectrum, and discuss the diffraction spectra of the
  other members of the family. Apart from the trivial Bragg peaks at
  the origin, they have purely singular continuous
  diffraction.\vspace{2ex}
\end{abstract}

Despite various open questions on details, the theory of substitutions
with pure point spectrum is fairly well developed. In particular, for
any given substitution on a finite alphabet, pure pointedness of the
spectrum can be decided algorithmically; see \cite{AGL} and references
therein. For substitutions of constant length, the situation is even
better due to Dekking's result \cite{D} and its recent extension by
Bartlett \cite{B}. In general, however, the understanding of
substitutions of non-constant length with mixed spectrum is still
rudimentary. Recent work has indicated that progress in this direction
is easier in the geometric setting of tiling spaces with natural tile
sizes; see \cite{BG,BFGR} and references therein. We adopt this point
of view here, too.

In this contribution, we consider the family of primitive substitution
rules on the binary alphabet $\{0,1\}$ defined by
\[
   \varrho_{m}^{}: \quad 0\mapsto 01^m\, ,\; 1\mapsto 0\, ,
   \quad
   \text{with } m\in\mathbb{N}. 
\]
The substitution matrix is $M_{m}=\left(\begin{smallmatrix}1 & 1\\ m &
  0 \end{smallmatrix}\right)$ with eigenvalues $\lambda_{m}^{\pm}
=\frac{1}{2} \bigl(1\pm\sqrt{4m+1}\,\bigr)$, which are the roots of
$\lambda^2-\lambda-m=0$. The frequency-normalised Perron--Frobenius
(PF) eigenvector is $(1,\lambda^{+}_{m}-1)^{t}/\lambda^{+}_{m}$, whose
entries are the relative frequencies of the two letters. The
corresponding left eigenvector reads $(\lambda^{+}_{m},1)$, which is
our choice of the interval lengths for the corresponding geometric
inflation rule.  Up to scale, this is the unique choice to obtain a
self-similar \emph{inflation tiling} of the line from
$\varrho^{}_{m}$; see \cite[Ch.~4]{TAO} for background.

From now on, we will mainly work with the tiling system on the real
line.  Note that $\varrho^{}_{1}$ defines the ubiquitous
\emph{Fibonacci tiling}, which is well-known to have pure point
spectrum, both in the diffraction and in the dynamical sense
\cite{Q,TAO,BL}. For $m=2$, we obtain a system that is equivalent to
the \emph{period doubling chain}, as can be seen by choosing
$a\,\widehat{=}\, 0$ and $b\,\widehat{=}\,11$ which establishes a
mutual local derivability (MLD) rule; see \cite[Secs.~4.5.1 and
9.4.4]{TAO} for details on the period doubling system. More generally,
$\lambda^{+}_{m}$ is an integer if and only if $4m+1$ is a
square. This precisely happens for $m=\ell(\ell+1)$ with
$\ell\in\mathbb{N}$, giving $\lambda^{+}_{m}=\ell+1$ and
$\lambda^{-}_{m}=-\ell$.  Similar to the $m=2$ case, any of these
systems can be recoded as a constant length substitution, via
$a\,\widehat{=}\, 0$ and $b\,\widehat{=}\,1^{\ell+1}$. The induced
substitution is $a\mapsto ab^{\ell}$, $b\mapsto a^{\ell+1}$, which has
a coincidence in the first position.  Consequently, one has pure point
spectrum by Dekking's criterion \cite{D}.  In all remaining cases, the
PF eigenvalue is irrational, but fails to be a Pisot--Vijayaraghavan
(PV) number, which means that none of the corresponding tilings can
have non-trivial point spectrum \cite{Boris,BFGR}.  In particular, the
only Bragg peak is the trivial one at $k=0$. So far, we have the
following result.

\begin{theorem}\label{thm1}
  Consider the inflation tiling defined by\/ $\varrho^{}_{m}$.  For\/
  $m=1$ and\/ $m=\ell (\ell+1)$ with\/ $\ell \in \mathbb{N}$, the
  tiling has pure point diffraction, which can be calculated with the
  projection method. For all remaining cases, the pure point part of
  the diffraction consists of the trivial Bragg peak at\/ $0$, while
  the remainder of the diffraction is of continuous type.
  \hfill $\Box$
\end{theorem}

\begin{figure}
\centerline{\includegraphics[width=0.8\textwidth]{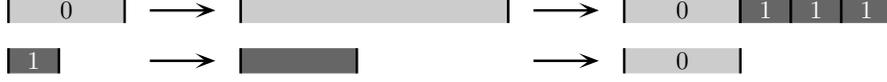}}
\caption{\label{fig:infl}Illustration of the self-similar inflation
  rule derived from $\varrho^{}_{3}$.}
\end{figure}

The first example with continuous component emerges for $m=3$, where
$\lambda^{\pm}_{3}=\frac{1}{2}\bigl(1\pm\sqrt{13}\,\bigr)$. Let us
discuss this case in some more detail (the other non-PV cases are
fairly analogous). The concrete inflation rule for this case is
illustrated in Figure~\ref{fig:infl}.  With $\varrho:=\varrho^{}_{3}$,
consider the bi-infinite fixed point $w$ of $\varrho^{2}$ with legal
seed $0|0$ obtained as
\[
  0|0 \,\stackrel{\varrho^{2}}{\longmapsto}\,  w^{(1)}=0111000|0111000
   \,\stackrel{\varrho^{2}}{\longmapsto}\,  \dots
   \,\stackrel{\varrho^{2}}{\longmapsto}\, w^{(i)}\; 
   \xrightarrow{\;i\to\infty\;}\; w=\varrho^{2}(w); 
\]
see \cite{Q,TAO} for background. The corresponding tiling, via the
left endpoints of the two types of intervals, leads to a Delone set
\[
   \varLambda^{w} \; =\; \bigl\{ \ldots, -1\!-\!3\lambda, 
     -3\lambda, -2\lambda,
       -\lambda, 0, \lambda, 1\!+\!\lambda, 
     2\!+\!\lambda, 3\!+\!\lambda,
       3\!+\!2\lambda,\ldots\bigr\} \; \subset\; \mathbb{Z}[\lambda],
\]
where $\lambda=\lambda^{+}_{3}$ from now on. Defining the (compact)
hull
$\mathbb{Y} := \overline{\{ t + \varLambda^{w} \mid t\in\mathbb{R}
  \}}^{\mathrm{LT}}$
with the closure being taken in the local topology (LT), we obtain a
topological dynamical system $(\mathbb{Y},\mathbb{R})$ under the
translation action of $\mathbb{R}$. This system is \emph{strictly
  ergodic}, which means it is \emph{minimal} (because $\mathbb{Y}$
coincides with the local indistinguishability (LI) class of
$\varLambda^{w}$) and \emph{uniquely ergodic} (since there is a unique
invariant probability measure $\mu^{}_{\mathbb{Y}}$, which is induced
by the uniformly existing patch frequencies for the corresponding
cylinder sets).

One important consequence is that every $\varLambda\in\mathbb{Y}$ has
the same autocorrelation and the same diffraction measure as
$\varLambda^{w}$, which are thus also called the autocorrelation and
diffraction of the dynamical system. More generally, given any
$\varLambda\in\mathbb{Y}$, we consider the weighted Dirac comb
$\omega := \sum_{x\in\varLambda} u(x) \delta_{x}$ with general complex
weights $u(x)\in\{u^{}_{0},u^{}_{1}\}$ according to the interval
type. Then, the corresponding autocorrelation $\gamma^{}_{u}$ is of
the form
\[
   \gamma^{}_{u} \, =  \!\sum_{z\in\varLambda-\varLambda} \!
   \eta^{}_{u}(z)\, \delta_{z}\quad
   \text{with}\quad 
   \eta^{}_{u}(z) \, = \lim_{r\to\infty} \,
   \frac{1}{2 \hspace{1pt} r} \!
   \sum_{y,y+z\in \varLambda_{r}}\!\! \overline{u(y)}\, u(y+z),
\]
where $ \varLambda_{r}:= \varLambda_{}\cap [-r,r]$.

Since we do not have any projection method at our disposal for the
further analysis, other tools are needed. It has recently been shown
in \cite{BG,BFGR} that the \emph{pair correlation functions} of a
primitive inflation tiling satisfy a set of \emph{exact
  renormalisation relations} that help to unravel the spectral type of
the diffraction. Partitioning
$\varLambda = \varLambda^{(0)}\dot{\cup} \varLambda^{(1)}$ into the
two point types, one can define the pair correlation functions as
\[
    \nu^{}_{ij}(z) \; = \; \lim_{r\to\infty}
    \frac{\mathrm{card} \bigl(\varLambda^{(i)}_{r} \cap 
    (\varLambda^{(j)}_{r} - z)\bigr)}{\mathrm{card} (\varLambda_{r})} 
    \; = \; \frac{1}{\mathrm{dens} (\varLambda)}\, \lim_{r\to\infty}
    \frac{\mathrm{card} \bigl(\varLambda^{(i)}_{r} \cap 
    (\varLambda^{(j)}_{r} - z)\bigr)}{2 \hspace{1pt} r} 
\]
for $i,j\in\{0,1\}$. These functions are well-defined for any
$z\in\mathbb{R}$, are non-negative, and satisfy the symmetry relations
$\nu^{}_{ij}(z) = \nu^{}_{ji}(-z)$. Moreover, $\nu^{}_{ij}(z)>0$ if
and only if $z\in\varLambda^{(j)}-\varLambda^{(i)}$, which is a
consequence of strict ergodicity.

The autocorrelation coefficients $\eta^{}_{u}(z)$ can now be expressed as
\[
   \eta^{}_{u}(z) \; = \; \mathrm{dens}(\varLambda) \!\! 
   \sum_{i,j \in\{0,1\}}\!\!
   \overline{u^{}_{i}} \; 
   \nu^{}_{ij}(z) \, u^{}_{j}  
\]
and similar expressions will later emerge for the diffraction measure
$\widehat{\gamma^{}_{u}}$. As follows from \cite{BG,BFGR}, one has the
following result.

\begin{theorem}\label{thm2}
The pair correlations functions\/ $\nu^{}_{ij}$ satisfy the linear
renormalisation equations
\begin{eqnarray*}
\nu^{}_{00}(z) & \!\! = \!\! & \frac{1}{\lambda} \,\Bigl(
  \nu^{}_{00}\bigl(\tfrac{z}{\lambda}\bigr) +
  \nu^{}_{01}\bigl(\tfrac{z}{\lambda}\bigr) +
  \nu^{}_{10}\bigl(\tfrac{z}{\lambda}\bigr) +
  \nu^{}_{11}\bigl(\tfrac{z}{\lambda}\bigr)\Bigr), \\
\nu^{}_{01}(z) & \!\! = \!\! & \frac{1}{\lambda} \,\Bigl( 
  \nu^{}_{00}\bigl(\tfrac{z-\lambda}{\lambda}\bigr) +
  \nu^{}_{00}\bigl(\tfrac{z-1-\lambda}{\lambda}\bigr) +
  \nu^{}_{00}\bigl(\tfrac{z-2-\lambda}{\lambda}\bigr) + 
  \nu^{}_{10}\bigl(\tfrac{z-\lambda}{\lambda}\bigr) +
  \nu^{}_{10}\bigl(\tfrac{z-1-\lambda}{\lambda}\bigr) +
  \nu^{}_{10}\bigl(\tfrac{z-2-\lambda}{\lambda}\bigr)\Bigr),\\
\nu^{}_{10}(z) & \!\! = \!\! &  \frac{1}{\lambda} \,\Bigl(
  \nu^{}_{00}\bigl(\tfrac{z+\lambda}{\lambda}\bigr) +
  \nu^{}_{00}\bigl(\tfrac{z+1+\lambda}{\lambda}\bigr) +
  \nu^{}_{00}\bigl(\tfrac{z+2+\lambda}{\lambda}\bigr) + 
  \nu^{}_{01}\bigl(\tfrac{z+\lambda}{\lambda}\bigr) +
  \nu^{}_{01}\bigl(\tfrac{z+1+\lambda}{\lambda}\bigr) +
  \nu^{}_{01}\bigl(\tfrac{z+2+\lambda}{\lambda}\bigr)\Bigr),\\
\nu^{}_{11}(z) & \!\! = \!\! &  \frac{1}{\lambda} \,\Bigl(
  3\, \nu^{}_{00}\bigl(\tfrac{z}{\lambda}\bigr) +
  2\, \nu^{}_{00}\bigl(\tfrac{z+1}{\lambda}\bigr) +
  2\, \nu^{}_{00}\bigl(\tfrac{z-1}{\lambda}\bigr) +
  \nu^{}_{00}\bigl(\tfrac{z+2}{\lambda}\bigr) +
  \nu^{}_{00}\bigl(\tfrac{z-2}{\lambda}\bigr) \Bigr) .
\end{eqnarray*}
Subject to the condition that the support of each $\nu^{}_{ij}$ is 
$\varLambda^{(j)}-\varLambda^{(i)}$, the solution space of this
infinite system of linear equations is one-dimensional.\hfill $\Box$ 
\end{theorem}

\begin{figure}
\centerline{\includegraphics[width=0.75\textwidth]{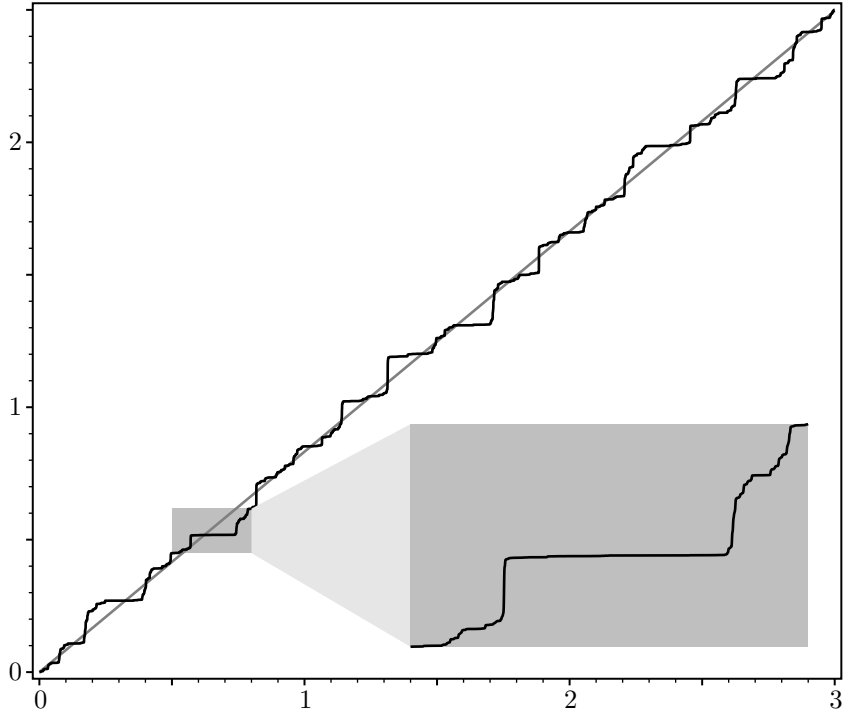}}
\caption{\label{fig:dist}Sketch of the distribution function $F$ for
  the singular continuous diffraction measure discussed in the
  text. The average slope of $F$ is $\eta^{}_{u}(0)$, 
  which is indicated by a straight line.}
\end{figure}

Via $\varUpsilon_{ij} := \sum_{z \in S_{ij}} \nu^{}_{ij} (z) \,
\delta^{}_{z}$, the pair correlation functions are turned into pure
point measures. The autocorrelation measure $\gamma^{}_{u}$ can now be
written as $ \gamma^{}_{u} (\mathcal{E}) = \mathrm{dens} (\varLambda)
\sum_{i,j \in \{ 0,1\} } \overline{u^{}_{i}} \; \varUpsilon^{}_{ij}
(\mathcal{E}) \, u^{}_{j}$, where $\mathcal{E}\subset\mathbb{R}$ is
any bounded Borel set. Taking the Fourier transform, one obtains the
diffraction measure as
\[
   \widehat{\gamma^{}_{u}} (\mathcal{E}) \, = \, 
   \mathrm{dens} (\varLambda)
   \sum_{i,j \in \{ 0,1\} } \overline{u^{}_{i}} \; \widehat{\varUpsilon}^{}_{ij}
   (\mathcal{E}) \, u^{}_{j},
\]
where the Fourier transform of each term can be shown to exist
\cite{BG}. Now, the renormalisation relations from Theorem~\ref{thm2}
induce measure-valued relations for the
$\widehat{\varUpsilon}^{}_{ij}$, which have to be satisfied for each
part of their Lebesgue decomposition separately. An analysis of the
asymptotic behaviour of the absolutely continuous components shows
that the only contribution compatible with local integrability of the
Radon--Nikodym densities and the translation boundedness of
$\widehat{\gamma^{}_{u}}$ is the trivial one, which means that no
absolutely continuous component is possible \cite{BFGR}. Consequently,
one has the following result.

\begin{theorem}
The diffraction measure\/ $\widehat{\gamma^{}_{u}}$, which is the same
for all\/ $\varLambda\in\mathbb{Y}$, has the pure point part\/
$\big\lvert \tfrac{2\lambda-1}{13}u^{}_{0} + \tfrac{7-\lambda}{13}
u^{}_{1} \big\rvert^{2}\, \delta^{}_{0}$.  The remainder of\/
$\widehat{\gamma^{}_{u}}$ is singular continuous.\hfill $\Box$
\end{theorem}

To give an impression of the singular continuous part, let us choose
$u^{}_{0}=1-\lambda$ and $u^{}_{1}=1$. With this choice, the Bragg
peak at $0$ is extinct, so $\widehat{\gamma^{}_{u}}$ is purely
singular continuous, with $\eta^{}_{u}(0)=(6\lambda-3)/13\approx
0.832$. Consequently, the \emph{distribution function} $F$ defined by
$F(x):=\widehat{\gamma^{}_{u}}\bigl([0,x]\bigr)$ is continuous, with
average slope given by $\eta^{}_{u}(0)$; see Figure~\ref{fig:dist} for
an illustration. Note that $F$ is strictly increasing, despite the
appearance of `flat' regions which resemble plateaux.

Let us close by commenting on the other non-Pisot members of our
inflation tiling family. So, let $m$ be any integer such that the pure
point part of the diffraction, according to Theorem~\ref{thm1}, is
trivial. Then, with our choice of interval lengths from the beginning,
the Bragg peak at the origin has intensity
\[
   I^{}_{0} \, = \, \big\lvert \mathrm{dens} (\varLambda)\,
  (u \cdot v^{}_{\text{PF}})\big\rvert^{2} \, = \, 
  \frac{\,\big\lvert u^{}_{0} + (\lambda^{+}_{m}-1) u^{}_{1} 
   \big\rvert^{2}}{4m+1}  ,
\]
where $\mathrm{dens}
(\varLambda)=\frac{\lambda^{+}_{m}}{2\lambda^{+}_{m}-1} =
\frac{\lambda^{+}_{m}+2m}{4m+1}$ and $v^{}_{\text{PF}}$ is the
PF frequency vector from above.\smallskip

For each such inflation tiling, the pair correlation functions are
again well-defined, and satisfy a set of exact linear renormalisation
relations in analogy to Theorem~\ref{thm2}. Completing the
corresponding analysis on the Fourier side, our analysis indicates
that we can never have an absolutely continuous component. A numerical
calculation of the distribution function analogous to $F$ above
produces graphs that are very similar to the one shown in
Figure~\ref{fig:dist}.

\end{document}